# Resistivity scaling in 1111 iron-pnictide superconductors

E. Arushanov[1,2]*, S. Levcenko[2+], G. Fuchs[1], and S.-L. Drechsler[1]

[1]Leibniz-Institut für Festkörper- und Werkstoffforschung Dresden - IFW Dresden, Helmholtzstr. 20, D-01069 Dresden, Germany

[2]Institute of Applied Physics, Academy of Sciences of Moldova, MD2028 Chisinau, Moldova

**Abstract.**

We show that the zero field normal-state resistivity above $T_c$ for various levels of electron doping - both for $LaO_{1-x}F_xFeAs$ (La-1111) and $SmO_{1-x}F_xFeAs$ (Sm-1111) members of the 1111-iron-pnictide superconductor family - can be scaled in a broad temperature range from 20 to 300 K onto single curves for underdoped La-1111 (x=0.05-0.075), for optimally and overdoped La-1111 (x=0.1-0.2) and for underdoped Sm-1111 (x=0.06-0.1) compounds. The scaling was performed using the energy scale $\Delta$, the resistivity $\rho_\Delta$ and the residual resistivity $\rho_0$ as scaling parameters as well as by applying a recently proposed model-independent scaling method (H. G. Luo, Y. H. Su, and T. Xiang, *Phys. Rev.* B **77**, 014529 (2008)). The scaling parameters have been calculated and the compositional variation of $\Delta$ has been determined. The observed scaling behaviour for $\rho(T)$ is interpreted as an indication of a common mechanism which dominates the scattering of the charge carriers in underdoped La-1111, in optimally and overdoped La-1111 and in underdoped Sm-1111 compounds..

Electronic mail: * arushanov@hotmail.com

[+] Levcenco@yandex.ru

# 1. INTRODUCTION

Recently, transition-metal oxypnictides composed by an alternate stacking of $Ln_2O_2$ layers and $T_2Pn_2$ layers (Ln:La, Pr, Ce, Sm, Nd; T: Fe, Co, Ni, Ru; Pn: P or As) have been identified as novel high-$T_c$ materials. [1-4] Their superconductivity discovered two years ago with a transition temperature $T_c$ of as high as 26 K in $LaO_{1-x}F_xFeAs$ [1] which reached in the mean time even about 57 K in corresponding rare earth substituents provided almost immediately a deep impact in contemporary condensed-matter physics, since this new superconductor does not belong to at least directly any known categories of '' old high-temperature superconductors'' such as copper-oxides (cuprates), fullerenes, and $MgB_2$.[5] Although the crystal structure of the transition-metal oxypnictides is different from that of cuprate superconductors, both compounds share similarities. (i) Superconductivity and quasi-metallic behaviour in the normal state are governed by two-dimensional networks with a relatively low concentration of charge carriers: the $CuO_2$ and the FeAs planes, respectively.
(ii) As in the cuprates and other unconventional superconductors, the parent undoped and the underdoped systems are antiferromagnets. Doping the latter systems for instance with fluorine suppresses both the magnetic order and the structural distortion in favour of superconductivity. Hence, like for high-$T_c$ copper oxides, the superconducting regime in these iron-based materials occurs in close proximity to a long-range-ordered antiferromagnetic ground state. Anyhow, there are important differences between the FeAs based materials and cuprates.[6] Thus, the parent compounds in the cuprates are antiferromagnetic charge-transfer-insulators remaining insulators also above the Néel-Temperature $T_N$, whereas LaOFeAs is an antiferromagnetic "spin-density wave" metal with parts of the former Fermi surface which remains ungapped also below $T_N$. The bad metallic behaviour is mainly caused by chemical reasons, leading to an almost filled mainly Fe-3$d$ state derived band complex which exhibits an almost compensated character with several (two) both electron ($e$) and hole ($h$) Fermi surface sheets (bands)). Their ($e$-$h$) interband nesting properties are believed to be responsible for the antiferromagnetism related properties. In the phenomenological description presented below this multiband character doesn't play a role but it should be keep in mind in interpreting the obtained results microscopically or inversely our results may provide useful constraints for a future detailed multiband description with a large amount of involved still unknown microscopic parameters.

The electronic phase diagrams of $LaO_{1-x}F_xFeAs$ (La-1111) [7] and $SmO_{1-x}F_xFeAs$ (Sm-1111)[7, 8] oxypnictide superconductors in the normal state based on the analysis of the electrical resistivity $\rho$ have been reported for a wide range of doping. The data give evidence for unusual normal state properties in these new materials.[7, 8] As a function of doping $x$ of $LaO_{1-x}F_xFeAs$, $\rho(T)$ shows a transition from pseudogap to Fermi liquid-like behaviour. The pseudogap signatures become stronger in Sm-1111.[7] The electronic phase diagram of the FeAs-superconductors yields a resemblance to the generic phase diagram of cuprate superconductors in the vicinity of a quantum critical point.[7] The Hall coefficient ($R_H$) reported for the $SmO_{1-x}F_xFeAs$ (x = 0–0.2)[8] is negative and decreases with increasing x, indicating that F doping leads to an increase in carrier concentration, and the dominating carriers are electrons.

The high-field magnetotransport properties of the oxypnictide superconductors $RFeAsO_{1-x}F_x$ ($R$=La, Nd)[9, 10] and $SmFeAsO_{1-\delta}$[10] have been studied by several authors. In particular, Jaroszinki et al.[10] came to the conclusion that $LaFeAsO_{0.89}F_{0.11}$ behaves as an intermediate-$T_c$ superconductor like $MgB_2$ in which thermal fluctuations of vortices do not significantly affect the H-T diagram at variance with the large extent that they do in the layered cuprates. However, the situation is different for the "high-$T_c$"- oxypnictides containing rare earth elements such as $SmFeAsO_{1-\delta}$, and $NdFeAs(O,F)$, which exhibit a larger mass anisotropy, enhanced thermal fluctuations, and for which the Ginzburg parameter becomes comparable to that of YBCO. Thus, the series of oxypnictide superconductors bridges a conceptual gap between conventional superconductors and the high-temperature cuprates.

An insight into the normal state might be in principle helpful also for the understanding of the superconductivity mechanism itself.[11] The evolution of the normal transport properties of the high-$T_c$ superconductor cuprates (HTSC) and that of other unconventional superconductors with temperature and doping still retains some features that are not yet understood. Clues which might help to solve some of the remaining questions can be obtained from the temperature dependence scaling of the normal state transport properties.[12] The scaling analysis of experimental data in HTSC is a simple, but powerful tool in elucidating the underlying physics without invoking a specific model. In the normal state, if the pseudogap is a predominant energy scale controlling low energy excitations, then the low-temperature behaviour of any measured physical quantity should satisfy a doping independent scaling law.[13] Recently, it has been shown that the temperature-dependent resistivity $\rho(T)$ and the Hall coefficient $R_H(T)$ of $YBa_2Cu_3O_x$, $SmBa_2Cu_3O_x$ and $La_{2-x}Sr_xCuO_4$ can be scaled.[12-19]

In this work, we report scaling of the normal-state transport properties of $LaO_{1-x}F_xFeAs$ and $SmO_{1-x}F_xFeAs$ in a broad temperature range from $T_c$ up to 300 K. The electron concentration in both samples has been systematically changed by varying the F content $x$.[7] These new scaling observations can be expanded towards other iron oxypnictides and might provide some important guidelines for the theoretical models proposed to describe beside the superconductivity in future also the normal-state properties of the new high-$T_c$ superconductors.

## II. RESULTS AND DISCUSSIONS

A commonly adopted approach in the scaling analysis is to assume that by normalizing both the measurement quantity $F(T)$ and the temperature by the corresponding values at a sample dependent characteristic temperature $T^*$, then all the experimental data should fall onto a single curve described by the scaling function.[13] In fact, the $\rho(T)$ and $R_H(T)$ of $YBa_2Cu_3O_x$, $SmBa_2Cu_3O_x$ and $La_{2-x}Sr_xCuO_4$ have been successfully scaled.[12, 14-19] The temperature is rescaled with a temperature $T_0$ defined as the temperature above which $\rho(T)$ shows a linear dependence[14,15], or with $T_\Delta$ where $\Delta$ is estimated on the basis of an analysis of the nonlinear part of $\rho(T)$[12,16], or with $T_H$, at which $R_H(T)$ changes from an essentially temperature-independent ($T > T_H$) to a rapidly increasing behaviour ($T < T_H$)[17,18] or with $T_R$, determined from the analysis of the exponential temperature dependence of $R_H$.[19] The latter is correlated to an activation energy $E_R$ that can be interpreted as the difference between the Fermi level and the saddle-point position observed in the electronic band structure of a $CuO_2$ plane.

Luo et al.[13] had proposed a model-independent scaling method to study the physical properties of high-temperature superconductors in the normal state. It was shown that the resistivity, the Hall coefficient, the magnetic susceptibility, the spin-lattice relaxation rate, and the thermoelectric power exhibit good scaling behavior, controlled purely by the pseudogap energy scale in the normal state.

The measured physical quantity $F(T)$ satisfies a simple, but rather general scaling law

$$F(T) = A_i f(\frac{T}{B_i}) + C_i , \qquad (1)$$

where

$A_i = \alpha_i/\alpha_s$, $B_i = \Delta_i/\Delta_s$, and $C_i = \beta_i - A_i\beta_s$ .

The parameters $\alpha$, $\beta$ and $\Delta$ are all doping dependent quantities, but being temperature independent. $\Delta$ is a characteristic energy scale of the system. The subscript $i$ is a sample index relatively to which the physical quantity $F$ as a function of $T$ is measured experimentally. The subscript $s$ is a reference sample index. For the reference sample, the scaling function is the measurement curve itself: $F(T) = f(T)$.

The relative scaling parameters $A_i$, $B_i$ and $C_i$ can be determined by minimizing the total deviation of the scaling functions:

$$\delta f = \sum_{i \leq j}^{N} \sum_{k}^{N_k} \left[ f_i(T_k) - f_j(T_k) \right]^2 ,$$

where

$$f_i(T) = \frac{1}{A_i} \left[ F_i(B_i T) - C_i \right] .$$

The minimization of $\delta f$ can be done, for example, using the standard subroutine given in Ref. 22 or the simulated annealing (SA) algorithm [23-25].

We apply scaling methods reported in Refs. 12, 13, 15, 16 to analyze the experimental data of new high-$T_c$ iron-pnictide superconductors in the normal state. We will first analyze the scaling behavior of the resistivity for $LaO_{1-x}F_x FeAs$. Our scaling analysis is based on the experimental data recently presented in Refs. 7 and 9. It should be mentioned that the character of the normal-state temperature dependence of $LaO_{1-x}F_x FeAs$ in underdoped (x= 0.05; 0.06 and 0.075) and optimally and overdoped compositions (x=0.1; 0.125; 0.15 and 0.2) is quite different. The former shows a linear $\rho(T)$-dependence near room temperature, a minimum of resistivity at about 60 K, and a low-temperature upturn. The latter shows linearity with temperature in the range of about 250-300 K[7], superlinear dependence at lower temperatures ($\rho \propto T^n$, where n= 2 (2.16) at temperature below 150 K[7] (220 K[20], x=0.1)) and a tendency to saturation at the lowest temperatures.[7,10,20]

We assume the following expression for the scaling function of $\rho$

$$\rho(T) = \rho_0 + c\, T \exp(-\Delta/T) \tag{2}$$

(where $\rho_0$ is the residual resistivity and $\Delta$ defines the energy scale controlling the linear and superlinear behaviour) which has been applied to $YBa_2Cu_3O_x$ and $Y_{1-x}Pr_xBa_2Cu_3O_x$ cuprate superconductors [12, 15, 16] and could be used to fit the temperature dependence of the resistivity of our samples, too. It should be noted that a similar expression $\rho(T) = \alpha T/\Delta\, \exp(\Delta/T)$ has been used by Luo et al.[13] to analyse the scaling behaviour of the in-plane resistivity of $YBa_2Cu_3O_x$, $Bi_2 Sr_2 CaCu_2O_{6+x}$ and $Bi_2 Sr_2 Ca_2Cu_3O_{8+x}$.

Eq. (2) proposed by Moshchalkov[15] to describe 1D transport has been successfully applied to underdoped cuprates in order to describe a universal superlinear resistivity $\rho(T)$.[12, 15, 16] We have found that Eq. (2) permits to get a good fit for $\rho(T)$ of $LaO_{1-x}F_x FeAs$ samples with x from 0.1 to 0.2 in a broad temperature region (Fig. 1a) and satisfactorily fit the underdoped samples (Fig.1b). This allows us to determine the absolute values of the scaling parameters for our samples (Table 1).

The obtained values of our fitting parameters $\rho_0$ and $\Delta$ are presented in Table 1. It can be seen that the value of $\Delta$ show a slight increase with x in the optimal doped - overdoped regimes (x=0.1-0.2) similarly to that observed for overdoped $Bi_2Sr_2Ca_2Cu_3O_{8+x}$.[24] and some decrease of $\Delta$ with decreasing x from x=0.1-0.2 to x=0.05-0.075 in underdoped samples. The obtained values of $\Delta$ between 10 and 19 meV are in satisfactory agreement with $\Delta$ = 15-20 meV reported by Sato et al.[5] which was estimated for $LaO_{0.93}F_{0.07}FeAs$ on the base of high-resolution photoemission spectroscopy.

The residual resistivity $\rho_0$ has been also extracted as a fitting parameter. The strong increase of $\rho_0$ with decreasing x can be considered as an indication that reducing the F content apart from lowering the electron charge carrier density also induces surprisingly an appreciable disorder. It is worth mentioning that the obtained values of $\rho_0$ are in agreement with those estimated as a fitting parameter of the low-temperature $\rho(T)$ dependence (below 200 K) by the equation $\rho(T) = \rho_0 + aT^n$. The values of $n$ are about 1.7 and 2 for samples with x=0.05-0.075 and 0.1-0.2, respectively. Similar results (n=2-2.16 at x=0.1) have been reported in Refs. 7 and 25. The zero-field data show that the samples studied are rather poor conductors having relatively low residual resistivity ratios $\rho_{290}/\rho_0$ (of 1.7 to 15 for x=0.075-0.1, see Table 1) compared to pure normal metals where $\rho_{290}/\rho_0 \sim 1000$. This agrees with previous findings on all superconducting high-$T_c$ cuprates and indicates their relative impure state.[15]

In Fig. 2 the scaled $\rho(T)$ curve is plotted for the $LaO_{1-x}F_xFeAs$ samples with $0.1 \leq x \leq 0.2$. According to Refs. 16 and 19 the temperature is rescaled with $\Delta$ and the resistivity is plotted as $(\rho - \rho_0) / (\rho_\Delta - \rho_0)$ where $\rho_\Delta$ is the resistivity at $T = \Delta$. All the $\rho(T)$ curves collapse onto one universal curve. The curve is linear with $T$ for $T/\Delta > 0.8$ (Fig. 2). At lower temperatures the $\rho(T)$ curves deviate from linearity and a superlinear $\rho(T)$ behavior sets in. Similar behaviour has been reported for underdoped YBCO cuprate superconductors.[12, 15, 16]

It has been shown by Luo et al[13] that the scaling curves could be obtained by using a model-independent scaling method. The method has been successfully applied to analyse the scaling behaviour of the normal-state transport properties of different cuprate superconductors such as Y123, Bi2212, Bi2223, La214, La-Bi221 etc.[13] We successfully applied the method both for underdoped (x=0.05- 0.075) as well as optimal and overdoped samples (x=0.-0.2). Samples with x=0.06 and 0.1 have been chosen as the reference samples for the former and the latter set of samples. As a result, an excellent scaling behaviour has been achieved for both underdoped (Fig.3), optimal, and overdoped samples (Fig.4). The scaling parameters *A, B* and *C* are given in Table 1. Taking into account that the value of *B* is equal to the ratio of the Δ values of the sample under study and the reference sample, the values of $\Delta_i$ have been calculated for all samples studied. The obtained values for $\Delta_i$ are in good agreement with those for Δ previously estimated (see Table 1). It can be seen that the values of Δ slightly increase with decreasing (increasing) x in the underdoped (overdoped) case.

We also analyzed the scaling behavior of *ρ(T)* of $SmO_{1-x}F_xFeAs$ for superconducting compositions (x=0.06; 0.08 and 0.1) presented in Ref. 7. The *ρ(T)* normal-state temperature dependence shows a decrease of the resistivity with decreasing temperature at temperatures studied and a linear temperature dependence of *ρ(T)* in the range of 250-300K (x=0.6 and 0.8) and between 50-60 and 120-130K (x=0.08 and x=0.1). [7]

A scaling function of *ρ(T)* for $SmO_{1-x}F_xFeAs$ is not available. Eq. (2) used to analyse the scaling of the *ρ(T)* dependence of La-1111 don't show a satisfactory fit in the case of Sm-1111. Using the method proposed by Luo *et al.* [13] we have studied the scaling of the normal-state transport properties of Sm-1111 and estimated the relative scaling parameters. The sample with the composition 0.1 has been used as a reference sample. A good scaling behaviour has been achieved for the samples studied (Fig. 5). The scaling parameters *A, B* and *C* are given in Table 1. It is worth to be mentioned that Liu *et al.* [8] pointed out on a possible scaling behavior between the Hall angle and temperature: cot $(\Theta_H)= \rho/ R_H$ vs $T^{1.5}$ for all $SmO_{1-x}F_x FeAs$ samples studied with x between 0 and 0.2, however, the scaling parameters have not been determined and the scaling curve has not been plotted.

It should be mentioned that both *Δ* and the characteristic temperature *T*\* shows a similar compositional dependence.[15] In overdoped $La_{2-x}Sr_x CuO_4$ (x=0.22-0.34)[18] and in $Bi_2Sr_2 CaCu_2O_{8+\delta}$ (δ= 0.25- 0.28) [24] the values of Δ and *T*\* are almost constant [18] or show a slight increase with x.[24] In the underdoped regime both values show a strong increase with decreasing x.[15, 18]

The observed compositional dependence of Δ in our overdoped La-1111 iron-pnictide superconductors is similar to those reported in Refs. 18, 24 for $La_{2-x}Sr_x CuO_4$ [18] and $Bi_2Sr_2 CaCu_2O_{8+\delta}$ superconductors. [24] In the underdoped regime the value of Δ in both iron-pnictide superconductors studied show a slight compositional dependence and being lower than in the overdoped samples. This is in contrast to the behavior reported for cuprate superconductors. [12-18] It is worth also mentioning that the As deficiency leads to a decrease of the value of Δ from 164 (x=0.1, δ=0) to about 120K (x=0.1, δ ≈ 0.05 to 0.1, see Table 1).

Following Refs. 14-16 we point out that the scaling behaviour of $\rho(T)$ observed in LaO$_{1-x}$F$_x$FeAs (x=0.05-0.075 and 0.1-0.2) and SmO$_{1-x}$F$_x$FeAs (x=0.06- 0.1) suggests that the transport properties in the studied samples in the above mentioned interval of $x$ are dominated by the same scattering mechanisms. Taking into account some similarity of the observed scaling behaviour with that reported for cuprate superconductors [16] we could assume that the dominant scattering mechanisms are probably also of magnetic origin. However, in contrast to cuprates the resistivity exhibit a different scaling behaviour for underdoped and overdoped LaO$_{1-x}$F$_x$FeAs samples which could indicate some difference in their character of the scattering mechanisms.

Our observation of a doping independent scaling law in the normal state of pnictide superconductors studied here similar to that observed in cuprates [13] could be probably used as indication that a gap like feature is the reason for a predominant energy scale which controls the low energy excitations. The possible existence of a pseudogap in La-1111 and Sm-1111 compounds has been reported by Sato et al. [5], Liu et al. [26], Jia et al. [27] and Ou et al. [28] on the base of high resolution photoemission measurements. However, its origin as to whether it can be caused by (i) local SDW fluctuation or (ii) CDW(charge density wave)/BOW(bond order wave) fluctuations supported by a strong enough electron-boson coupling and special intersite Coulomb interactions needs further experimental and theoretical studies. [28] A clear signature of a superconducting gap opening below $T_c$ was observed in the far-infrared reflectance spectra of LaFeAsO$_{0.9}$F$_{0.1-\delta}$. [29] Fluctuations of a mesoscopically heterogenous phase separation state like the well-known stripe state in cuprates might involve features of all components (SDW, CDW and BOW) mentioned above. A tendency to electronically driven phase separations has been reported for some hole doped 122 pnictides superconductors as well as the electron doped 122 systems with Co doping at Fe-sites.

The characteristic energy scale on which the scattering occurs in LaO$_{1-x}$F$_x$FeAs (x=0.05-0.075 and 0.1-0.2) and SmO$_{1-x}$F$_x$FeAs (x=0.06-0.1) which is determined by the temperature $\Delta$ shows a slight variation with x both in underdoped and overdoped regimes (Table 1). In addition, the values of $\Delta$ in underdoped LaO$_{1-x}$F$_x$FeAs samples are even lower than those in overdoped samples. The behavior observed in our iron oxypnictide samples is very different from that in cuprate superconductors. Our results are in accord with reported 'pseudo gap-like features observed in recent high resolution angle-integrated photoemission spectroscopy measurements.[29] These authors have found a doping independent behavior of the spectral weight suppression in SmO$_{1-x}$F$_x$FeAs (x = 0, 0.12, 0.15, 0.2). A large "pseudogap" of 80 meV has been reported for the highest measured temperature, whose origin is currently unclear, and debatably, could be extrinsic. A smaller gap of 10meV becomes observable below 100K, which is likely an intrinsic pseudogap in the normal state, and might be more or less closely related to one of the observed gaps in the superconducting state.

IV. CONCLUSIONS

We have demonstrated that the zero-field normal-state resistivity for various levels of doping for La-1111 and Sm-1111 iron-pnictide superconductors can be scaled onto single curves for underdoped La-1111 (x=0.05-0.075), for optimally and overdoped La-1111 (x=0.1-0.2) and for underdoped Sm-1111 (x=0.06-0.1) compounds. One of the two scaling methods used here is based on the assumption that the scaling behaviour is approximately given by $\rho(T) = \rho_0 + c\, T \exp(-\Delta/T)$. It was found that an energy scale $\Delta$, the resistivity $\rho_\Delta$ and the residual resistivity $\rho_0$ are suitable scaling parameters for $LaO_{1-x}F_xFeAs$. The second one, a recently proposed model-independent scaling method[13] to study the physical properties of high-temperature superconductors in the normal state has been applied to analyse the scaling behaviour of both the La-1111 and the Sm-1111 systems as well. An excellent scaling of the normal-state resistivity for both material groups has been observed. The results obtained by two different methods are in good agreement with respect to the presence of a characteristic energy $\Delta$ which resembles the pseudo gap feature in cuprates. The observed scaling behaviour for $\rho(T)$ is interpreted as an indication of a single mechanism which dominates the scattering of the charge carriers in underdoped La-1111 (x=0.05-0.075), in optimally and overdoped La-1111 (x=0.1-0.2) and in underdoped Sm-1111 (x=0.06-0.1) compounds. The physical nature of this gap like feature remains unclear at present. Among other possibilities at least three scenarios are worth to be studied theoretically in more detail. (i) A long-range Coulomb interaction disorder related feature. (ii) A pairing mechanism related pseudo-gap as discussed for the cuprates and (iii) a band structure related feature from a band slightly below the Fermi energy. Also a corresponding study of other less two dimensional iron pnictides of the 122 and the 111 families as well as of related 11 iron selenides or tellurides is of considerable interest and might be helpful to discriminate between the proposed scenarios.

ACKNOWLEDGMENT

One of us (EA) would like to thank the DFG for financial support, G. Behr[†] and B. Büchner for support and V. Gvozdikov and A. Moebius (SLD) for discussions concerning the

Coulomb interaction and the pseudogap problem possibly related to 2D systems close to localization.

**Figure captions**

**Fig.1.** Temperature dependence of the resistivity for (a) optimally and overdoped $LaO_{1-x}F_xFeAs$ (x=0.1 and x=0.2, respectively) and (b) underdoped $LaO_{1-x}F_xFeAs$ (x=0.075). The solid lines represent a fit using Eq. 2.

**Fig.2.** Scaling analysis on the temperature dependence of the resistivity of various $LaO_{1-x}F_xFeAs_{1-\delta}$ systems ($0.1 \leq x \leq 0.2$) including also an arsenic-deficient sample. The temperature is rescaled with $\Delta$ (an energy scale) and the resistivity is given by $\frac{\rho - \rho_0}{\rho_\Delta - \rho_0}$ in which the extrapolated residual resistivity $\rho_0$ has been subtracted and $\rho_\Delta$ is the resistivity at $T= \Delta$.

**Fig.3.** The scaled resistivity for underdoped $LaO_{1-x}F_xFeAs$ ($0.05 \leq x \leq 0.075$).

**Fig.4.** The scaled resistivity for optimally and overdoped $LaO_{1-x}F_xFeAs$ ($0.1 \leq x \leq 0.2$) including a arsenic-deficient sample.

**Fig.5.** The scaled resistivity for underdoped $SmO_{1-x}F_xFeAs$ ($0.04 \leq x \leq 0.1$).

**Table 1.** Residual resistivity $\rho_0$, the ratio $\rho_{290}/\rho_0$, $T_c$ as well as the scaling parameters $\Delta$, $A$, $B$ and $C$.

| Samples | $\rho_0$ ($10^{-5}\Omega$m) | $\rho_{290}/\rho_0$ | $2\Delta$ (K) | A | B | $2\Delta_i$ [a] (K) | C ($10^{-5}\Omega$m) | $T_c$ [b] (K) |
|---|---|---|---|---|---|---|---|---|
| **LaO$_{1-x}$F$_x$FeAs** | | | | | | | | |
| x=0.05 | 1.95 | 1.98 | 121 | 1.33 | 1.07 | 132 | 0 | 21 |
| x=0.06 | 1.48 | 2.05 | 124 | 1.0 | 1.0 | **124** | 0 | 20 |
| x=0.075 | 1.63 | 2.11 | 120 | 1.11 | 0.97 | 120 | 0 | 22 |
| x=0.1 | 0.210 | 10.76 | 164 | 1.0 | 1.0 | **164** | 0 | 26 |
| x=0.125 | 0.212 | 8.82 | 179 | 0.97 | 1.14 | 187 | 0.03 | 18 |
| x=0.15 | 0.138 | 15.14 | 219 | 1.45 | 1.32 | 217 | -0.15 | 10 |
| x=0.2 | 0.139 | 13.67 | 217 | 1.29 | 1.30 | 213 | -0.11 | 9 |
| **LaO$_{1-x}$F$_x$FeAs$_{1-\delta}$** (x=0.1, $\delta\approx$0.05 to 0.1) | 0.606 | 6.6 | 114 | 1.13 | 0.75 | 123 | 0.41 | 28.5 |
| **SmO$_{1-x}$F$_x$FeAs** | | | | | | | | |
| x=0.06 | 0.409 | 2.42 | 392 | 0.63 | 0.80 | 345 | 0.26 | 36 |
| x=0.08 | 0.266 | 3.73 | 388 | 0.79 | 0.72 | 310 | 0.04 | 45 |
| x=0.1 | 0.205 | 4.8 | 431 | 1.0 | 1.0 | **431** | 0 | 52 |

[a] The values for $2\Delta_i$ are calculated as $B \times 2\Delta_s$ where $2\Delta_s$ for the three reference samples are shown in bold in the column for $2\Delta_i$ ($2\Delta_s$ = 124 K, 164 K and 431 K for underdoped La-1111, optimally and overdoped La-1111 and underdoped Sm-1111, respectively)
[b] The $T_c$ values were taken from Ref. 28.

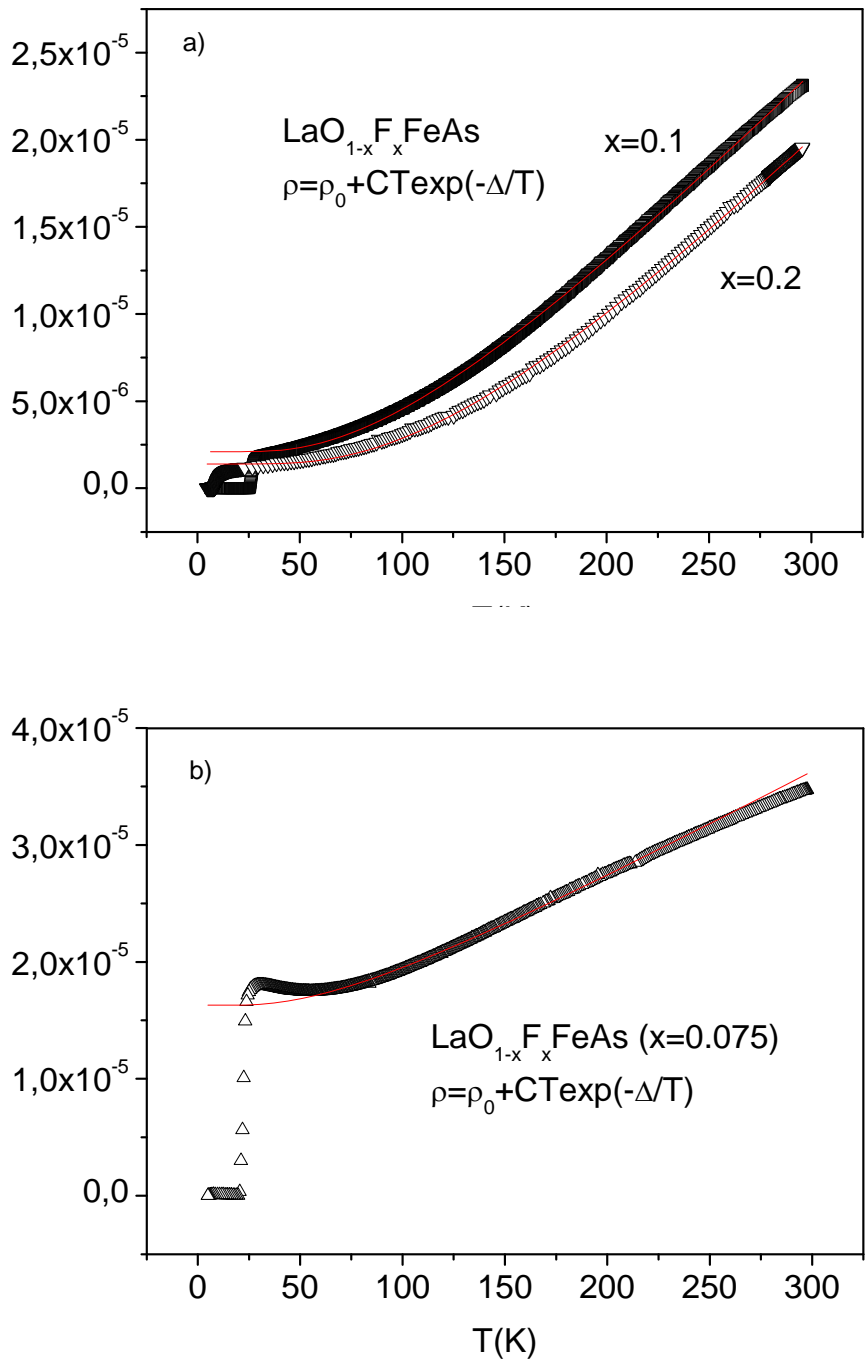

**Fig. 1.** Temperature dependence of the resistivity for (a) optimally and overdoped $LaO_{1-x}F_xFeAs$ (x=0.1 and x=0.2, respectively) and (b) underdoped $LaO_{1-x}F_xFeAs$ (x=0.075). The solid lines represent a fit using Eq. 2.

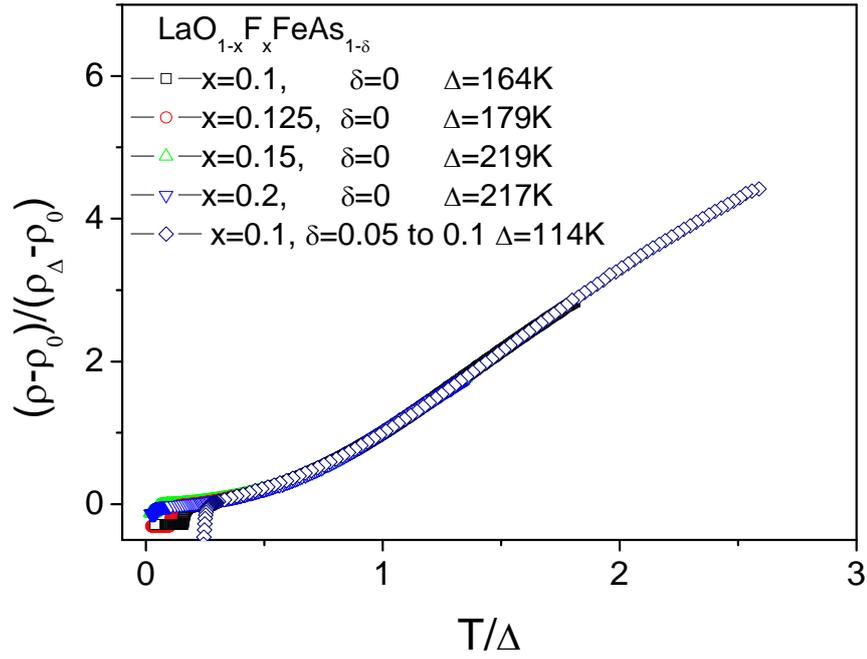

**Fig. 2.** Scaling analysis on the temperature dependence of the resistivity of various $LaO_{1-x}F_xFeAs_{1-\delta}$ samples ($0.1 \leq x \leq 0.2$) including also an arsenic-deficient sample. The temperature is rescaled with $\Delta$ (an energy scale) and the resistivity is given by $\frac{\rho - \rho_0}{\rho_\Delta - \rho_0}$ in which the extrapolated residual resistivity $\rho_0$ has been subtracted and $\rho_\Delta$ is the resistivity at $T = \Delta$.

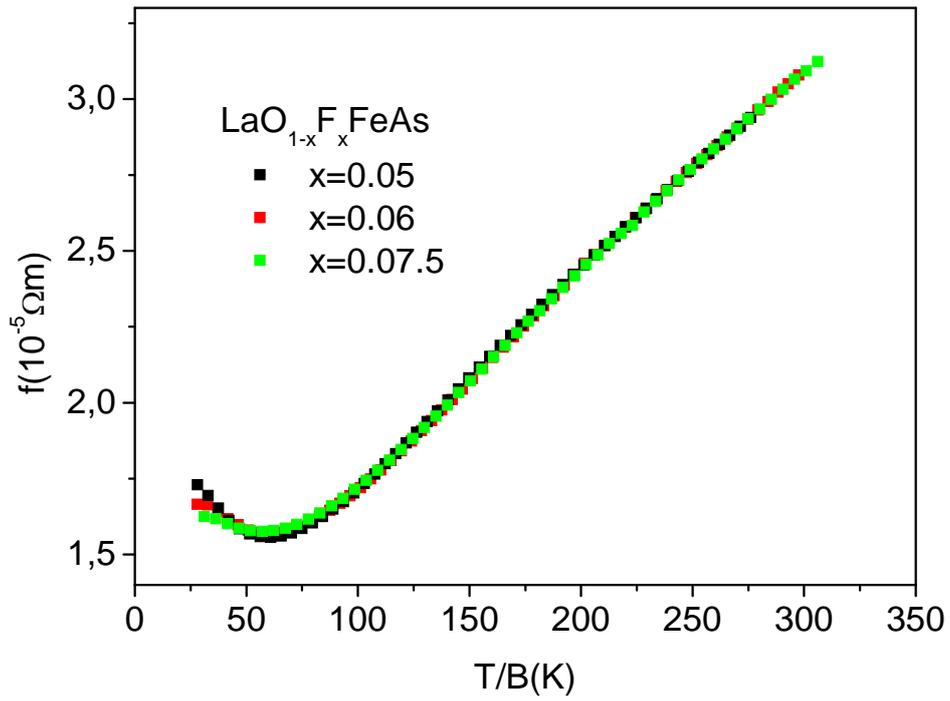

**Fig. 3**. The scaled resistivity for underdoped LaO$_{1-x}$F$_x$FeAs ($0.05 \leq x \leq 0.075$).

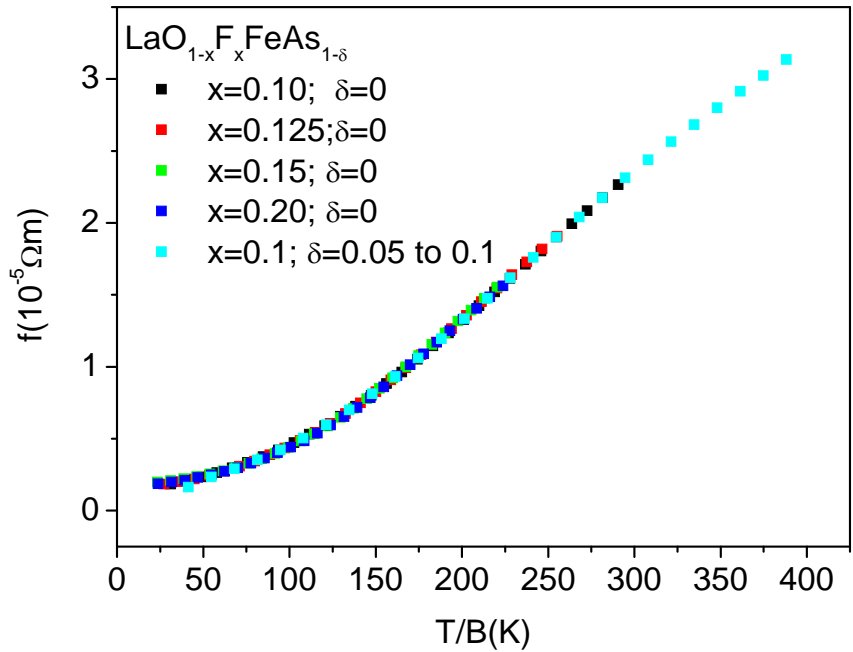

**Fig. 4.** The scaled resistivity for optimally and overdoped LaO$_{1-x}$F$_x$FeAs ($0.1 \leq x \leq 0.2$) including a arsenic-deficient sample.

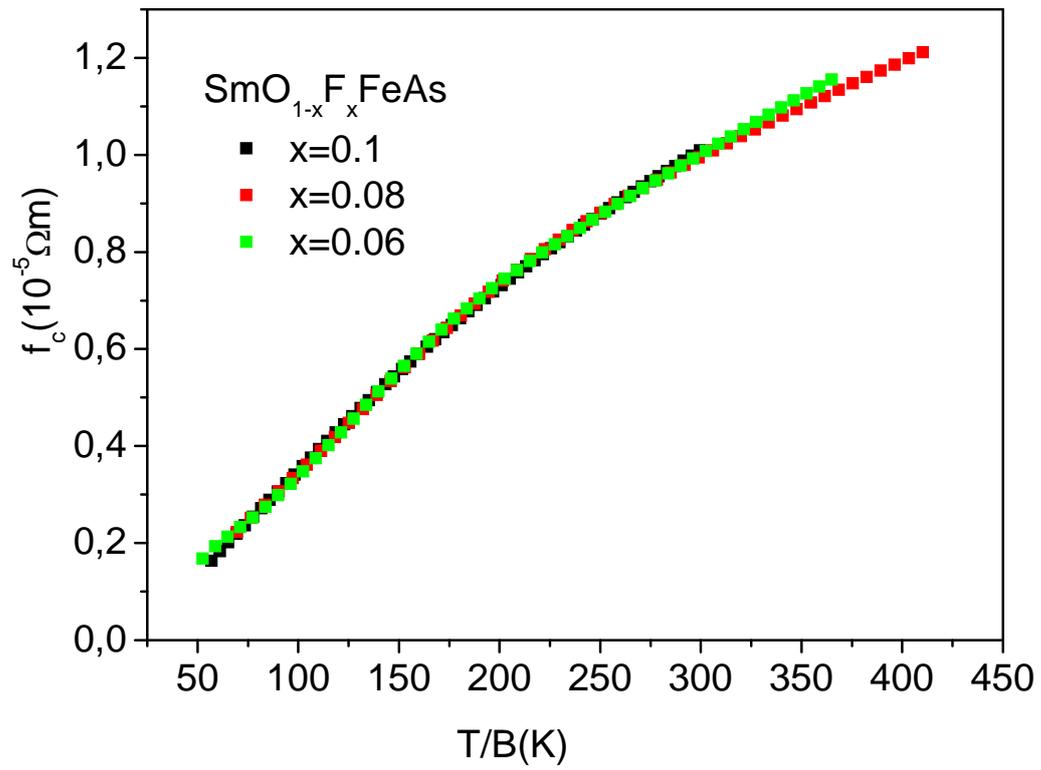

**Fig. 5.** The scaled resistivity for underdoped SmO$_{1-x}$F$_x$FeAs ($0.04 \leq x \leq 0.1$).